# Spin switching in electronic devices based on 2D assemblies of spin-crossover nanoparticles

*Julien Dugay\*, Mónica Giménez-Marqués, Tatiana Kozlova, Henny W Zandbergen, Eugenio Coronado\* and Herre S. J. van der Zant*

Dr. J. Dugay, T. Kozlova, Prof. H. W. Zandbergen, Prof. H. S. J. van der Zant

Kavli Institute of Nanoscience, Delft University of Technology, Lorentzweg 1, 2628 CJ Delft, the Netherlands

E-mail: julien.dugay@gmail.com

Dr. M. Giménez-Marqués, Prof. E. Coronado

Instituto de Ciencia Molecular, Unversidad de Valencia, Catedrático José Beltrán 2, 46980 Paterna, Spain

E-mail: eugenio.coronado@uv.es

Spin-crossover (SCO) materials form an intriguing class of compounds in which various external physical or chemical stimuli can change their ground spin state between a low-spin (LS) and a high-spin (HS) near room-temperature.[1–4] It has been a long standing dream to use this phenomenon in solid-state electronic devices.[5] Only recently several research groups succeeded in electrically addressing SCO materials based on Fe (II) coordination compounds at the nanoscopic[6–8] and microscopic[8–10] levels. This electric control of the spin has even been achieved in individual SCO molecules.[11,12]

The most investigated SCO systems are based on six-coordinated octahedral iron(II) complexes. In such systems, the spin state changes from the LS configuration, $S = 0$, to the HS one, $S = 2$; this SCO is accompanied by a change in the magnetic moment, the structure and color of the compound. Although the origin of this HS-LS transition is purely molecular, the SCO manifests cooperatively at the solid state and is strongly influenced by short/long range elastic interactions between neighboring SCO molecules. Such a cooperativity can result in a thermal hysteresis cycle, a property that can be used for memory applications in electronic devices.[5,13]

In recent years the size of the SCO solid has been reduced to the nanoscale, while preserving the memory effect near room temperature. In particular, nanoparticles (NPs) of ~10 nm[14,15] have been obtained, based on the polymeric one-dimensional (1D) compound ([Fe(Htrz)$_2$(trz)](BF$_4$)), (Htrz = 1,2,4-triazole),[16,17] which is one of the most promising SCO systems for designing electronic devices working at room temperature.[14] SCO NPs of other materials forming 3D networks have also been studied in this context, but in most cases they undergo a fast decrease of the hysteresis width upon size reduction, which has been accompanied by a higher residual high spin fraction and a downshift of the transition temperatures.[19]

Recently, the memory effect of the [Fe(Htrz)$_2$(trz)](BF$_4$) NP(s) has electrically been demonstrated,[6,8,9] showing their potential as switching elements of nanoscale devices. Nevertheless, opposite behaviour in the conductivity change during the SCO transition has been reported, which remains unclear mainly due the scare experimental studies of the effects of an electric field on the charge transport properties.[20,21] In particular, measurements on single NP devices (NP mean size of 10 nm) show an increase in the electric conductivity when the SCO occurs,[6] whereas the reverse situation is observed for an assembly of NPs (mean size 15 nm)[8] and for larger particles[8–10] (from 100 to 300 nm in diameter and from 250 to a couple of microns in length). Although it is tempting to attribute this difference to separate transport mechanisms operating in these devices, *i.e*, single-step tunnelling vs. multistep hopping, differences in the chemical and structural NP features as well as in the experimental conditions may not be neglected. Besides, the link between SCO NP morphologies and the memory effect features has not been established yet.

In this communication we study the transport properties of two-dimensional assemblies of ([Fe(Htrz)$_2$(trz)](BF$_4$)) SCO NPs with two different morphologies. The NPs have been synthesized made in a similar manner than in our previous study in which single NPs were measured.[6] The morphologies were characterized by HRTEM techniques revealing that these SCO systems possess a rod shape (with different lengths and diameters) and form a compact monolayer where their longest dimension is parallel to the sheet. By combining HRTEM techniques with the transport properties we correlate the shape of the thermal hysteresis loop in the conductivity, not only to the global assembling properties of the films, but also to the individual NPs morphology.

To allow a direct comparison between the TEM analysis and the electrical devices, we prepared free-standing ultrathin sheets of two batches of SCO NPs on top of holey carbon TEM grids in a similar manner as those used in the charge transport measurements discussed below. These sheets are composed of self-assembled monolayers of the SCO NPs, which were initially formed at a diethyeleneglycol/octane interface and subsequently transferred on holey carbon TEM grids (called hereafter the μ-contact printing method).[6] The NPs morphology as well as their global arrangement have been determined using high-angle annular dark field scanning transmission electron microscopy (STEM-HAADF), a technique which provides a high contrast of the NPs.

**Figure 1a and 1b** presents STEM-HAADF images of the self-assembled monolayers of NPs showing the two different morphologies. Both NPs display a rod-like shape. To the best of our knowledge only two studies reported X-ray measurements performed on single crystal of 1D polymeric compounds related to our system.[22,23] However, none of these studies permitted to relate the crystallographic data to the crystal orientation since indexation of the crystal faces was not included. We would like to mention that our attempts to gain structural information by TEM diffraction resulted unsuccessful due to beam damage of the NPs during the measurements (even though very soft conditions were employed by means of a low dose combined with imaging performed at 100 K). In this respect, because the single crystals obtained are also needle-shaped we assume for the next that the rod-like shape of the NPs is linked to the chain-like structure of the family of iron(II)-triazole complexes, with the chains aligned along their longest dimension (see **Figure S1** in the Supporting Information). **Figure 1** also evidences that both NPs are organized parallel to the sheet, which must be due to the strong capillary forces induced by the octane thin film during evaporation.

To get more detailed information of the cross sectional shapes of the NPs, free-standing sheets of vertically self-assembled monolayers were formed by means of a drop casting method on holey C TEM grids. At the locations of the holes in the C film, such a deposition method indeed promoted the formation of NP assemblies where the rod direction is parallel to the electron beam (see insets in **Figure 1a and b** display representative high magnification STAM-HAADF images).[24] The two viewing directions allow for an accurate determination of the sizes of the two NP batches (see size distribution analysis, **Figure 1c and 1d**). The SCO NP systems possess two different volumes ($V \approx 1960$ and $1240$ nm$^3$, respectively) corresponding to lengths of 25 nm and 44 nm along the rod direction and average diameters of 10 nm and 6 nm respectively. The two NP systems will be called hereafter **25/10 nm** and **44/6 nm** NPs respectively.

Planar finger-like shaped electrodes separated by about 50 nm were used for charge-transport measurements (see **Figure 2e** for an AFM image of the device before deposition). We transferred ultrathin sheets of SCO NPs on top of such electrodes. **Figure 2a and b** display respectively optical and AFM images of a typical device recorded after deposition (**25/10 nm** sized NPs; **sample A**). The pink areas in **Figure 2a** correspond to the ultrathin sheets of SCO

NPs that cover approximately 70% of the device. AFM images of the zoomed source-drain region (**Figures 2c and 2d**) confirm the 2D character of the NP sheet and are consistent with the previous STEM-HAADF observations. Importantly, we find that the sheets are not free-standing between the electrodes but in direct contact with the substrate (SiO$_2$), as seen from the significant height difference between the nanotrench and the electrodes (**Figures 2c and 2d**). We have recorded current-voltage I(*V*) characteristics continuously in time while slowly ramping the temperature to a target value. **Figure 3a** displays a 2D current map of *sample A* during heating up to 400 K.

**Figure 3b** shows thermal hysteresis loops in the conductance, obtained by subsequently cooling to room temperature, plotted at different bias voltages. We find that the width of the hysteresis loop is equal to 45 K, close to the value deduced from magnetization data performed on a powder sample (see **Figure S 3a** and **Table 2** in Supplementary Information). I(*V*) characteristics recorded within the thermal hysteresis region at 350 K in both the heating and cooling modes are reported in **Figure 3c** (in high- and low-conducting states, respectively). We identify these two states with the low and high spin states respectively, following the magnetization data on a powder of the same SCO NPs. The corresponding relative change of current in both spin states (I$_{LS}$ / I$_{HS}$) deduced from the I(*V*) curves is shown in **Figure 3d.** We restricted the bias voltage to the 10-15 V range, where the current values are above the noise level (~ 4 pA). The results in **Figure 3** indicate the absence of an applied-bias voltage effect on the global hysteresis features (relative change of electrical conductance as well as the hysteresis width and steepness). This means that the poorly conducting character of a few SCO NPs in parallel can be overcome by increasing the bias voltage without altering the characteristic properties of the spin transition.

It is worth noting that unlike the hysteresis loop deduced from magnetization data on sample powders, the one observed in the electrical conductance for NPs sheets remains open after the cooling mode has been finished (*i.e,* the conductance does not recover its initial value). This suggests that the thermal spin transition is not completely reversible (see **Figure 3b** at 300 K). Additionally, the conductance recorded during the very first heating cycle differs drastically from the successive thermal cycles; it is progressively shifted towards lower values until it falls below the detection limit. Based on studies that will be developed elsewhere, we ascribe the incomplete back-relaxation to the LS state, mainly to the progressive trapping of a fraction of $Fe^{II}$ species in the HS state during temperature cycles.

Subsequently, we measured the very first thermal hysteresis loop in the conductance of SCO NPs with the two different morphologies (reproduced on different devices fabricated and characterized under the same conditions). **Figure 4** presents the normalised and superimposed electrical characteristics of four samples: *sample A and B* correspond to SCO NPs of **25/10nm**, while *sample C and D* stand for devices with SCO NPs of **44/6 nm**. As shown in **Figure 4**, all devices exhibit a clear thermal hysteresis loop in the conductance. Moreover, the HS state (at high temperature) is systematically and significantly less conductive than the LS state, in agreement with previous studies involving assemblies of the same SCO systems.[8–10] Remarkably, this trend is reversed at the single particle level, where an increase of the electrical conductivity in the HS state was reported by us using the same NPs.[6] These observations lead to the conclusion that the electrodes distance determine how the conductance is influenced by the spin state. Importantly, we emphasize that other factors such as the deposition method, device preparation and sample preparation are not the responsible for the observed differences since in the present work these factors have been maintained the same as those employed in our previous work.[6]

Another interesting point concerns the influence of the SCO NPs morphologies on the shape of the hysteresis loop. A widening of the hysteresis loop as well as an increase of the transition temperatures ($T_{1/2}^{up}/T_{1/2}^{down}$) for samples **C** and **D (44/6 nm)** as compared to the ones from **A** and **B (25/10 nm)** are observed (see **Table 1)**. This behaviour is puzzling considering the opposite trend recorded on powder samples containing these SCO NPs (see hysteresis width values in **Figure S 3** and **Table 2,** Supplementary Information), commonly explained by the decrease of the number of intra-particle interactions with decreasing NP volume (*i.e,* loss of cooperativity).[17,21–23] This opposite behaviour may be related to the larger cooperativity expected in the longer nanorods (*i.e,* in NPs of **44/6 nm** as this length is directly connected with the length of the SCO chains), while in NPs of **25/10 nm** the number of Fe(II) centers is smaller along the Fe chains. Besides, a recent report showed a stiffness enhancement of SCO NPs when increasing their surface-to-volume ratio (*i.e,* for NPs of 44/6 nm the stiffness is expected to be the highest).[28] Thus, weak 2D inter-particle interactions for monolayer thick sheets could allow such stiffness/cooperativity enhancement to become observable, whereas the strong 3D inter-particle interactions in powders could mask its observation. The wider hysteresis loop in 2D and the upshift of the transition temperatures can be ascribed to the stiffness/cooperativity enhancement compensating the reduction of interactions due to the volume decrease. The long-range 3D interactions, on the other hand, could mask the morphology and surface effect competition, leading to a closening of the hysteresis loop.[29]

We also observed that the nanorods with the smallest volume (**44/6 nm**) do not exhibit a well-defined square-shaped hysteresis loop in the conductivity compared to the ones recorded for the samples with a larger volume (**25/10 nm**) (see **Figure 4**). This could reflect the importance of inter-chain interactions, which are expected to be more pronounced as the number of chains in the NP increases. Thus, the larger diameter of the NPs (**25/10 nm)** may explain their more abrupt spin transition.[17]

Contrastingly, we did not observe a clear size dependence of the hysteresis loop steepness in powder samples, especially for NPs measured after transport measurements (referred to as "old" in **Figure S 3**). This absence of a size effect when a 3D dimensionality is involved can be ascribed to a different packing of NPs in these powder samples, making the interpretation complicated.

A quite remarkable aspect of the data represented in **Figure 4** is the large change in the electrical conductance associated with the spin transition; it changes by about 2 orders and 1 order of magnitude for the 2D assemblies based on **25/10** and **44/6 nm** size SCO NPs, respectively. Remarkably, the values on the small volume NPs are larger than those reported in aggregates of rod-like shaped SCO triazole-based NPs of larger sizes organized at the macroscopic scale, for which changes of almost one order of magnitude have been reported.[9,10] At the nanoscale, measuring one single NP, we reported a small value (smaller than 3).[6] The high ratio in our assemblies may be related to the organization of the nanorods in the sheets, which are compactly aligned in 2D, thereby directing the electrical pathways to be preferentially aligned along the nanorod axis (i.e., along the iron chains referred as *b* axis in **Figure S1**). Along this direction, the relative change of the ligand-iron bond length during the LS to HS transition is the largest (anisotropic expansion of 6.3% along *b*, compared to 1% along *a* and 4% along *c*).[30,31] However, the nanoscale size of the particles must also play an important role, especially on the genuine lower conductance state, as micrometric nanorods placed by the dielectrophoresis technique between interdigitated electrodes with a very high degree of alignment only show a minor conductance change of 1.5.[8] At the present stage a detailed explanation of the strong change in the electrical conductance and its sharp size dependence remains unclear, and both further experimental investigations and theoretical support are needed.

In summary, sheets of self-assembled monolayers of SCO NPs formed at the air/liquid interface presenting a rod-like shape with two different morphologies (**25/10** and **44/6** nm) were electrically measured by contacting them on pre-patterned gold electrodes substrates. In a similar manner, we prepared free-standing monolayer sheets of both SCO NPs on holey carbon TEM grids to extract NP size distributions and their global arrangement within the sheets by STEM-HAADF technique.

For both NP morphologies, a thermal hysteresis loop in the electrical conductance near room temperature was observed, which was correlated to their morphologies and 2D organization. The most remarkable result was the observation of an unprecedented large change in the electrical conductance of the two spin states, increasing by up to two orders of magnitude in the 2D assemblies with **25/10 nm** sized SCO NPs. This value is significantly larger than that obtained in single-NP measurements ($\approx 3$). Although such large conductance changes are not fully understood, our results point at important contributions from its size/morphology-dependence at the nanoscale. In addition, we found that the HS state corresponds to a low-conductance state, which is consistent with previous studies performed on assemblies of SCO NP, albeit fabricated under different conditions. It is, however, in contrast to single-NP SCO studies with the same compound, which show that the HS state exhibits the high-conductance state. The experimental results ask for theoretical studies to gain more insights in the genuine role of the NP size and morphology, as well as the inter-particle interactions with the SCO transition.

**Experimental Section**

Electrodes were fabricated in one step by standard electron-beam lithography techniques. They consist of a 2-nm-thick Ti adhesive layer, with ~30 nm of Au on top. The substrate is made of intrinsic Si with a 285 nm thick silicon oxide layer grown on top to insulate the source and the drain electrodes from it. Before use, the samples with the electrodes were $O_2$ plasma cleaned during 15 minutes to remove organic residues.

25/10 and 44/6 nm sized SCO NPs were synthesized using a previously described procedure based on the reverse-micelle technique developed by some of us in 2007 with some modifications.[14] The size of the NPs was tuned by selecting the appropriate molar ratio [water] / [surfactant] (also known as $\omega_0$ parameter) in the micelles. In a first step two separate micellar solutions incorporating the different reactants were prepared: an aqueous solution of Fe(BF$_4$)$_2$.6H$_2$O (0.3 mL, 1 M) is added to a previously prepared solution of the surfactant dioctyl sulfosuccinate sodium salt, NaAOT (1.3 g for **25/10** and 1.0 g for **44/6**) in *n*-octane (10 mL). Similarly, an aqueous solution of Htrz (0.3 mL, 3 M) ligand is added to a solution of NaAOT (1.3 g for **25/10** and 1.0 g for **44/6**) in of *n*-octane (10 mL). Note, that the concentration of surfactant has been modified while the water content remains constant, so that the concentration of reagents is not altered. Then, the two-micellar solutions are stirred at room temperature until a thermodynamically stable microemulsion is formed (*ca.* 30 minutes). Once the microemulsions are stabilized, they are mixed at room temperature under Ar atmosphere and stirred to ensure the micellar exchange. A characteristic pink color appears immediately, indicating the occurrence of a nucleation process and the formation of the complex in an optically transparent suspension.

Powdered samples for magnetic measurements were obtained by precipitation upon addition of acetone to destabilize the microemulsion. SCO NPs were collected by centrifugation (12000 rpm, 10 minutes) after several washing cycles with portions of acetone (x3) and ethanol (x3) to remove excess of surfactant. Finally, the powdered samples were dried under vacuum for 2 h to remove excess of solvent.

The visualization of the SCO NPs was done using a FEI Titan microscope operating at 300 keV. Scanning transmission electron microscopy mode was used to diminish the damaging consequences of beam illumination. Using a high-angle annular dark field detector the contrast enhancement was achieved due to suppression of diffraction contrast while mass–thickness contrast leads to sufficient difference between high intensity NPs separated by almost no-intensity gaps.[32] The size distribution analysis was done by using the free ImageJ software.[33]

SCO NPs were deposited using a stamping technique as previously published.[6]

All transport measurements were performed in a cryostat probe station (Desert Cryogenics) with home-built low-noise electronics. The measurements were performed in ambient conditions under a small nitrogen flow to avoid the condensation of a water layer at the top of the electrical devices. Temperature dependence studies (300–400 K) were done using a built-in heater element with a Lakeshore temperature controller and measured locally by using a calibrated thermistor (TE-tech, MP-3011). The latter was stuck on top of a substrate identical to the ones used in this work (same silicon/SiO$_2$ thicknesses) and placed on the sample stage nearby the measured devices. The temperature was changed at a rate of 4 K·min$^{-1}$ in the heating mode by means of a resistor underneath the sample stage, while the cooling rate is controlled at high temperature only when the natural thermal dissipation is faster than 4 K·min$^{-1}$.

**Supporting Information**

Supporting Information is available from the Wiley Online Library or from the author.

**Acknowledgements**

We thank EU (Advanced ERC grants SPINMOL and Mols@Mols) and NWO/OCW**, the Spanish MINECO (grant MAT2011-22785) and the Generalidad Valenciana (PROMETEO and ISIC Programs of excellence) for financial support of this work. We thank Michele Buscema and Ferry Prins for experimental help and fruitful discussions.

Received:     ((will     be     filled     in     by     the     editorial     staff))
Revised:      ((will     be     filled     in     by     the     editorial     staff))
Published online: ((will be filled in by the editorial staff))

[1] P. Gütlich, H. A. Goodwin, J.-F. Létard, P. Guionneau, L. Goux-Capes, in *Spin Crossover Transit. Met. Compd. III*, Springer Berlin / Heidelberg, **2004**, pp. 1–19.
[2] K. S. Murray, C. J. Kepert, in *Spin Crossover Transit. Met. Compd. I* (Eds.: P. Gütlich, H.A. Goodwin), Springer Berlin Heidelberg, **2004**, pp. 195–228.
[3] J. A. Real, A. B. Gaspar, M. C. Muñoz, *Dalton Trans.* **2005**, 2062.
[4] P. Gamez, J. S. Costa, M. Quesada, G. Aromí, *Dalton Trans.* **2009**, 7845.
[5] O. Kahn, C. J. Martinez, *Science* **1998**, *279*, 44.
[6] F. Prins, M. Monrabal-Capilla, E. A. Osorio, E. Coronado, H. S. J. van der Zant, *Adv. Mater.* **2011**, *23*, 1545.
[7] C. Etrillard, V. Faramarzi, J.-F. Dayen, J.-F. Letard, B. Doudin, *Chem Commun* **2011**, *47*, 9663.
[8] A. Rotaru, J. Dugay, R. P. Tan, I. A. Guralskiy, L. Salmon, P. Demont, J. Carrey, G. Molnár, M. Respaud, A. Bousseksou, *Adv. Mater.* **2013**, *25*, 1745.
[9] A. Rotaru, I. A. Gural'skiy, G. Molnar, L. Salmon, P. Demont, A. Bousseksou, *Chem Commun* **2013**, *48*, 4163.
[10] C. Lefter, I. A. Gural'skiy, H. Peng, G. Molnár, L. Salmon, A. Rotaru, A. Bousseksou, P. Demont, *Phys. Status Solidi RRL – Rapid Res. Lett.* **2013**, *8*, 191.
[11] T. Miyamachi, M. Gruber, V. Davesne, M. Bowen, S. Boukari, L. Joly, F. Scheurer, G. Rogez, T. K. Yamada, P. Ohresser, E. Beaurepaire, W. Wulfhekel, *Nat. Commun.* **2012**, *3*, 938.
[12] T. G. Gopakumar, F. Matino, H. Naggert, A. Bannwarth, F. Tuczek, R. Berndt, *Angew. Chem. Int. Ed.* **2012**, *51*, 6262.
[13] J. Larionova, L. Salmon, Y. Guari, A. Tokarev, K. Molvinger, G. Molnár, A. Bousseksou, *Angew. Chem.* **2008**, *120*, 8360.
[14] E. Coronado, J. R. Galán-Mascarós, M. Monrabal-Capilla, J. García-Martínez, P. Pardo-Ibáñez, *Adv. Mater.* **2007**, *19*, 1359.
[15] J. R. Galán-Mascarós, E. Coronado, A. Forment-Aliaga, M. Monrabal-Capilla, E. Pinilla-Cienfuegos, M. Ceolin, *Inorg. Chem.* **2010**, *49*, 5706.
[16] G. Aromí, L. A. Barrios, O. Roubeau, P. Gamez, *Coord. Chem. Rev.* **2011**, *255*, 485.
[17] O. Roubeau, *Chem. – Eur. J.* **2012**, *18*, 15230.
[18] J. Kroeber, J.-P. Audiere, R. Claude, E. Codjovi, O. Kahn, J. G. Haasnoot, F. Groliere, C. Jay, A. and Bousseksou, *Chem. Mater.* **1994**, *6*, 1404.
[19] F. Volatron, L. Catala, E. Rivière, A. Gloter, O. Stéphan, T. Mallah, *Inorg. Chem.* **2008**, *47*, 6584.
[20] H. J. Shepherd, G. Molnár, W. Nicolazzi, L. Salmon, A. Bousseksou, *Eur. J. Inorg. Chem.* **2013**, *2013*, 653.
[21] E. Ruiz, *Phys. Chem. Chem. Phys.* **2014**, *16*, 14.
[22] M. M. Dîrtu, C. Neuhausen, A. D. Naik, A. Rotaru, L. Spinu, Y. Garcia, *Inorg. Chem.* **2010**, *49*, 5723.
[23] A. Grosjean, N. Daro, B. Kauffmann, A. Kaiba, J.-F. Létard, P. Guionneau, *Chem. Commun.* **2011**, *47*, 12382.
[24] D. Baranov, A. Fiore, M. van Huis, C. Giannini, A. Falqui, U. Lafont, H. Zandbergen, M. Zanella, R. Cingolani, L. Manna, *Nano Lett.* **2010**, *10*, 743.
[25] T. Forestier, A. Kaiba, S. Pechev, D. Denux, P. Guionneau, C. Etrillard, N. Daro, E. Freysz, J.-F. Létard, *Chem. – Eur. J.* **2009**, *15*, 6122.
[26] R. V. Martinez, J. Martínez, M. Chiesa, R. Garcia, E. Coronado, E. Pinilla-Cienfuegos, S. Tatay, *Adv. Mater.* **2010**, *22*, 588.
[27] A. Rotaru, F. Varret, A. Gindulescu, J. Linares, A. Stancu, J. F. Létard, T. Forestier, C. Etrillard, *Eur. Phys. J. B* **2011**, *84*, 439.
[28] G. Félix, W. Nicolazzi, M. Mikolasek, G. Molnár, A. Bousseksou, *Phys. Chem. Chem. Phys.* **2014**.
[29] A. Muraoka, K. Boukheddaden, J. Linarès, F. Varret, *Phys Rev B* **2011**, *84*, 054119.
[30] A. Grosjean, P. Négrier, P. Bordet, C. Etrillard, D. Mondieig, S. Pechev, E. Lebraud, J.-F. Létard, P. Guionneau, *Eur. J. Inorg. Chem.* **2013**, *2013*, 796.
[31] A. Urakawa, W. Van Beek, M. Monrabal-Capilla, J. R. Galán-Mascarós, L. Palin, M. Milanesio, *J. Phys. Chem. C* **2011**, *115*, 1323.
[32] M. Rudneva, T. Kozlova, H. W. Zandbergen, *Ultramicroscopy* **2013**, *134*, 155.
[33] W. S. Rasband, *Httprsbweb Nih Govij* **2008**.

**Figure 1. a-b)** Free-standing 2D sheets of SCO NPs prepared by µ-contact printing and imaged in STEM-HAADF mode. Insets: lateral self-assembly of NPs obtained by a drop-casting method. **c-d)** Distributions of NPs diameters (Φ: *9 ± 3 and 6 ± 2 nm*) and lengths (L: *25 ± 12 and 44 ± 11 nm*) for SCO NPs systems referred to as **25/10** (big volume) and **44/6** nm (small volume) in the main text.

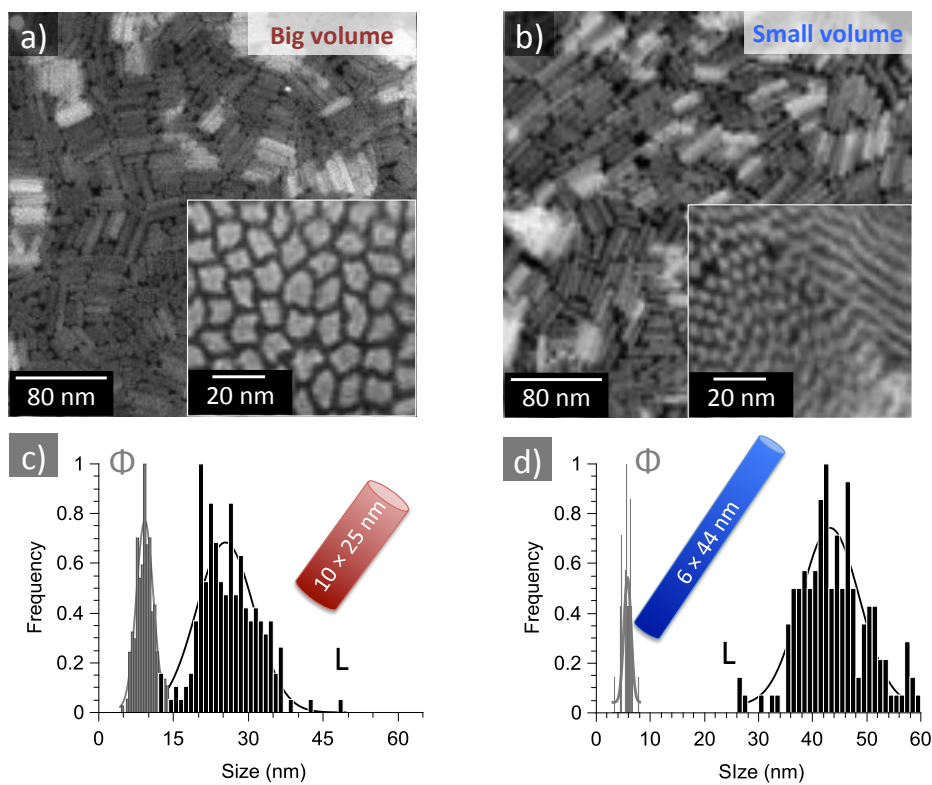

**Figure 2. a-b)** Optical and atomic force microscopy (AFM) images respectively of the electrical device recorded after deposition of **25/10 nm** sized SCO NPs. **c)** AFM zoom-in image of the source-drain region. **d)** Scaled up and cropped view of the nanotrench formed by the source-drain electrodes with a higher brightness revealing the NPs inside the gap. **e)** 3D AFM image of the fine pattern of the electrical device before deposition. Some panels show a height profile inset taken at the dashed line.

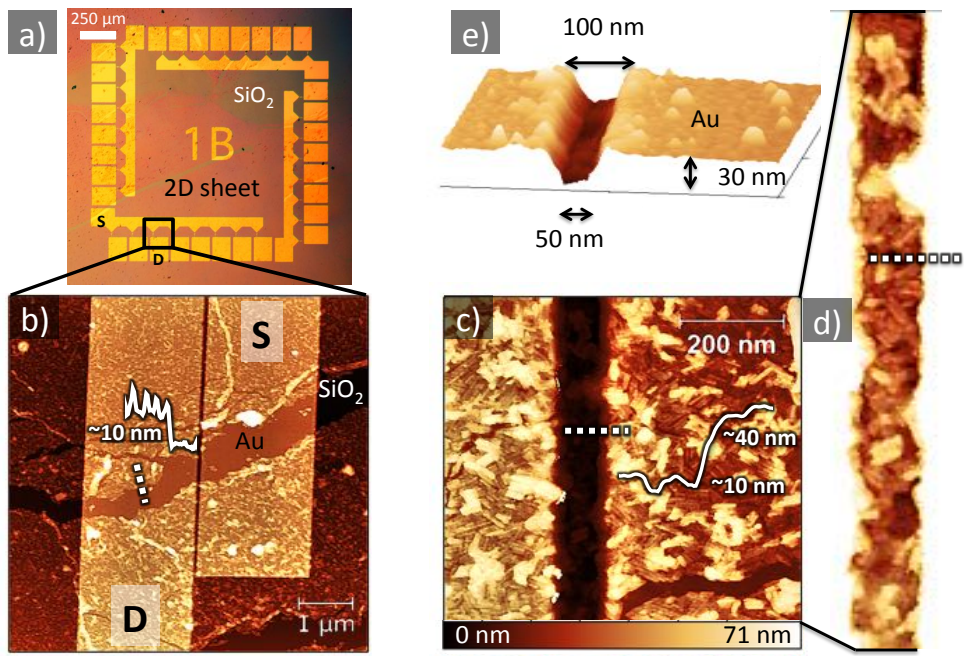

**Figure 3. SAMPLE A (25/10 nm): a)** Current-voltage I(*V*) characteristics recorded continuously while ramping the temperature from 300 to 400 K. **b)** Corresponding thermal variation of the electrical conductance plotted for different applied voltages (*V* = 15, 12.5, 10 V) in the heating and cooling modes. **c)** I(*V*) characteristics recorded within the thermal hysteresis region at 350 K in the heating (corresponding to the low-spin state) and cooling (high-spin state) modes. **d)** Relative conductance change associated with the spin transition ($I_{LS} / I_{HS}$) versus the bias voltage (*V*).

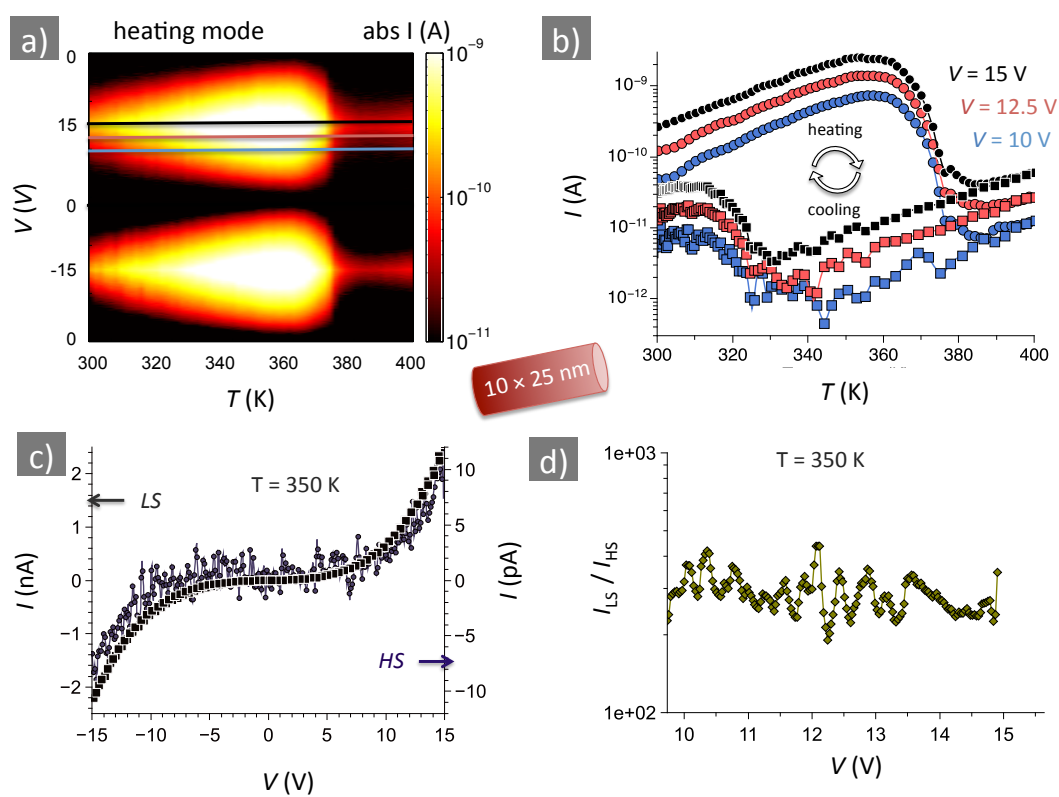

**Figure 4.** Electrical characterization of the SCO NPs 2D-sheets. Semi-log plot of the normalized conductance (G/G$_0$) measured as a function of temperature for NPs with a large (**25/10 nm**) and small (**44/6 nm**) volume. G$_0$ corresponds to the conductance value at room temperature (in the heating mode).

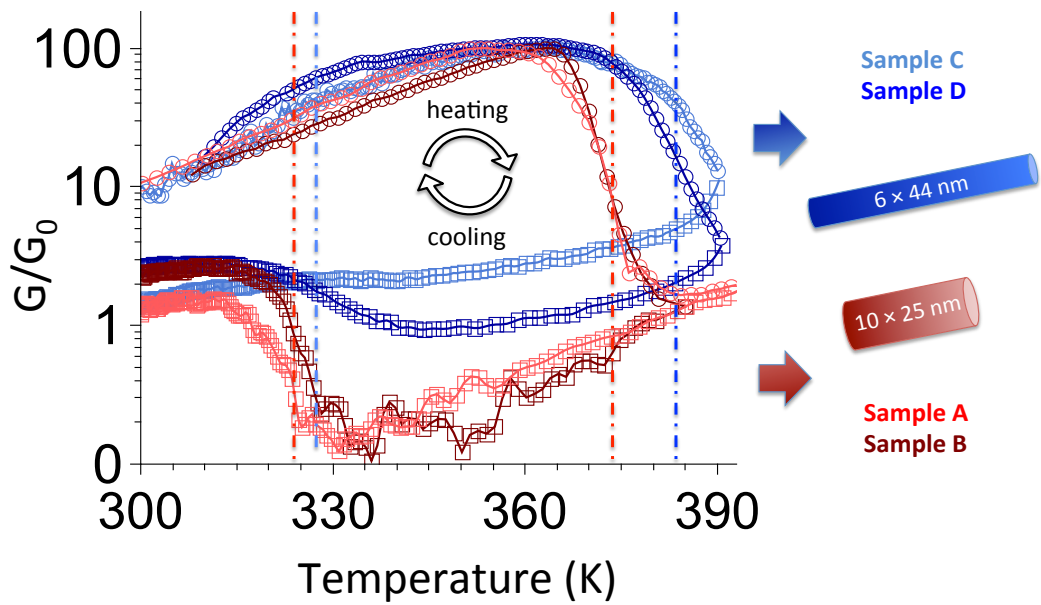

**Table 1**. Critical temperature and current values extracted from the hysteresis loop in the electrical conductance (see main text).

| Sample | Bias voltage max. | $T_{1/2}^{up}$ (K) [a] | $T_{1/2}^{down}$ (K) [a] | ΔT | Steepness | $I_{LS} / I_{HS}$ [b] |
|---|---|---|---|---|---|---|
| **1_A** | 15 V | 368 | 323 | 45 | "Squared" | 310 ± 79 |
| **1_B** | 4 V | 372 | 325 | 43 | "Squared" | 277 ± 97 |
| **2_C** | 10 V | 382 | 330 | 52 | "Smoothed" | 106 ± 3 |
| **2_D** | 10 V | 380 | 331 | 49 | "Smoothed" | 45 ± 2 |

[a] The error bar was estimated to be ± 2 K for the critical temperature (hence ± 4 K for the width), corresponding to the half width at the half maximum of the fit by a Gaussian function of the magnetic susceptibility derivative. [b] The error bar for each sample corresponds to the standard deviation of the data within the bias voltage range where $I_{LS} / I_{HS}$ is saturated (see main text).

2D organizations of triazole-based SCO NPs presenting different morphologies were electrically measured. A thermal hysteresis loop in the electrical conductance near room temperature was correlated with the NP morphologies and their 2D organization. An unprecedented large change in the electrical conductance of the two spin states is demonstrated, increasing by up to two orders of magnitude.

*Julien Dugay\*, Mónica Giménez-Marqués, Tatiana Kozlova, Henny W Zandbergen, Eugenio Coronado\* and Herre S. J. van der Zant*

Dr. J. Dugay, T. Kozlova, Prof. H. W. Zandbergen, Prof. H. S. J. van der Zant

Kavli Institute of Nanoscience, Delft University of Technology, Lorentzweg 1, 2628 CJ Delft, the Netherlands

E-mail: h.s.j.vanderzant@tudelft.nl

Dr. M. Giménez-Marqués, Prof. E. Coronado

Instituto de Ciencia Molecular, Unversidad de Valencia, Catedrático José Beltrán 2, 46980 Paterna, Spain

E-mail: eugenio.coronado@uv.es

**Spin switching in electronic devices based on 2D assemblies of spin-crossover nanoparticles**

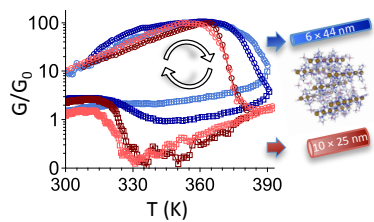

Supporting Information

**Spin switching in electronic devices based on 2D assemblies of spin-crossover nanoparticles**

*Julien Dugay\*, Mónica Giménez-Marqués, Tatiana Kozlova, Henny W Zandbergen, Eugenio Coronado\* and Herre S. J. van der Zant*

Dr. J. Dugay, T. Kozlova, Prof. H. W. Zandbergen, Prof. H. S. J. van der Zant

Kavli Institute of Nanoscience, Delft University of Technology, Lorentzweg 1, 2628 CJ Delft, the Netherlands

E-mail: h.s.j.vanderzant@tudelft.nl

Dr. M. Giménez-Marqués, Prof. E. Coronado

Instituto de Ciencia Molecular, Unversidad de Valencia, Catedrático José Beltrán 2, 46980 Paterna, Spain

E-mail: eugenio.coronado@uv.es

**Figure S1.** SCO NPs studied in this work are made of an organic shell (AOT molecules) surrounding an active core of the family of triazole-based polymeric compound [Fe$^{II}$(Htrz)$_2$(trz$^-$)](BF$_4^-$). The crystal structure is based on polymeric chains of Fe$^{II}$ ions situated along the *b* axis with the three bridging triazoles in alternating invert positions, two Htrz ligands and one trz$^-$ ligand (Pnma space group). Each [Fe(Htrz)$_2$(trz)]$_n$ chain is surrounded by six identical adjacent chains with two different Fe⋯Fe interchain distances resulting in a pseudo-hexagonal structure. The SCO produces a change of the intrachain Fe⋯Fe distance of 0.23 Å, with a variation of the *b* parameter of *ca.* 6.3 %, while the *a* and *c* parameters vary *ca.* 1.0 % and 4.0 %, respectively.

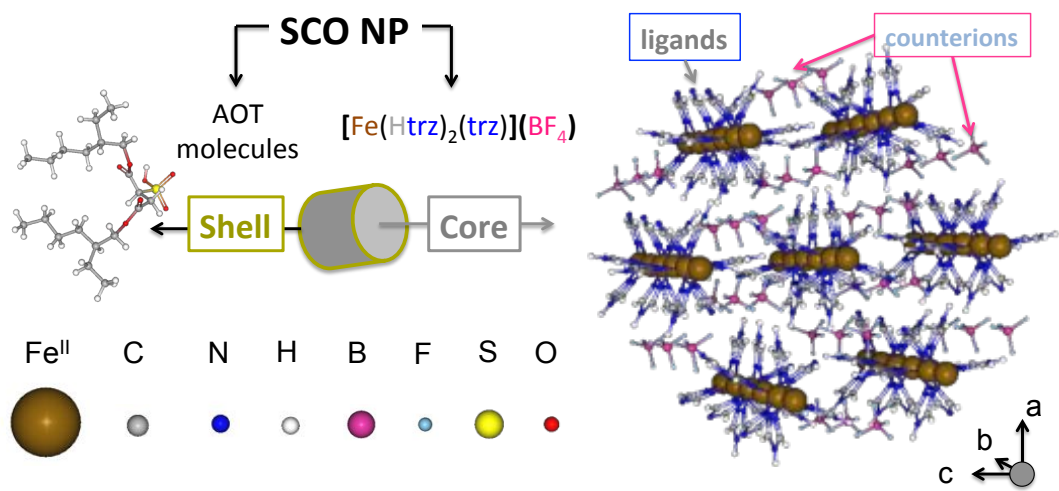

**Figure S2. a-b)** Low-magnification STEM-HAADF images of free-standing 2D sheets of laterally (respectively vertically) SCO NPs (referred as **25/10nm** size SCO NPs in the main text**)** self-assembled at a diethyeleneglycol/octane interface and fished after octane evaporation (respectively drop-casted) on holey carbon TEM grids. Insets show higher magnification of the free-standing sheets. **Magnified views** of this system are presented in **Figure 1a**.

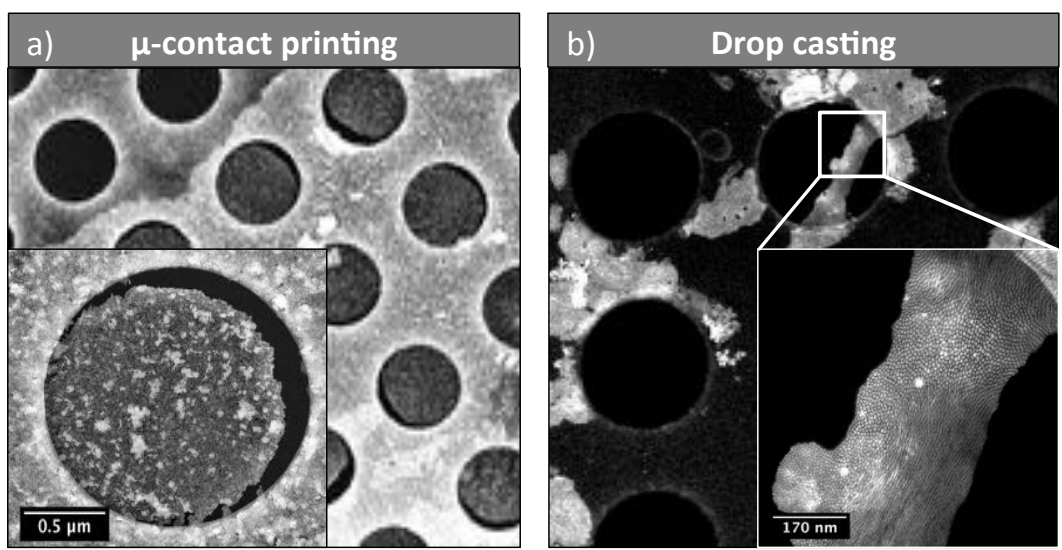

**Figure S 3.** Temperature dependence of the magnetic moment for powdered samples of **25/10 nm** (a) and **44/6nm** (b) size SCO NPs recorded in the temperature range 300-400 K under an applied magnetic field of 0.1 T (heating/cooling rate of 1 K.min$^{-1}$). Such a protocol was followed for a reference batch obtained just after synthesis (plotted with red and blue circles) and for a second batch of the same synthesis but from the same colloidal suspensions used for electrical measurements stored for several months (plotted with black squares).

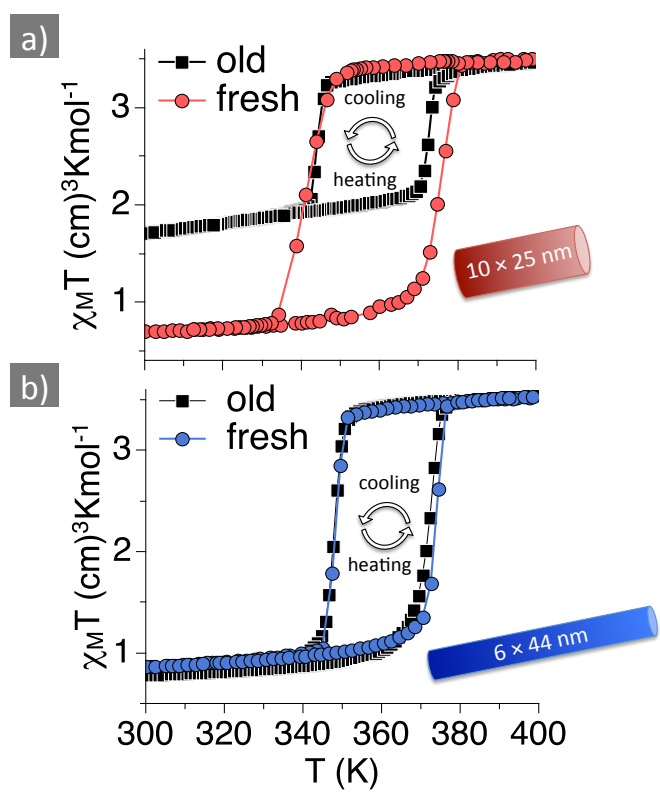

SCO NPs display cooperative magnetic and chromatic thermal–induced spin transition with a large hysteresis above room temperature. At high temperature the $\chi_M T$ value of both systems is typical of Fe$^{II}$ compounds in the HS state (*ca.* 3.5 emu.mol$^{-1}$.K), whereas at low temperatures different residual $\chi_M T$ values are observed corresponding to a large fraction (*26* and *19* % for fresh solutions of 25/10 and 44/6 nm size SCO NPs, respectively) of HS Fe$^{II}$. The subsequent warming mode reveals the occurrence of a thermal hysteresis loop of 34 and 25 K wide (transition temperatures: $T_{1/2}^{up}$ = 375 K and $T_{1/2}^{down}$ = 341 K, and $T_{1/2}^{up}$ = 372 K and $T_{1/2}^{down}$ = 344 K for fresh solutions of 25/10 and 44/6 nm size SCO NPs, respectively). Magnetic measurements of both NPs precipitated from the same suspensions after several months (old) reveal that 25/10 nm size SCO NPs present a much higher HS residual fraction (42 %) whereas 44/6 nm size NPs basically reproduce the spin transition.

**Table 2.** Features of the hysteresis loop extracted from magnetization data (see above).

| Sample | $T_{1/2}^{up}$ (K)[a] | $T_{1/2}^{down}$ (K)[a] | ΔT | High-spin fraction |
|---|---|---|---|---|
| **1_fresh** | 375 | 341 | 34 | 26 % |
| **1_old** | 372 | 344 | 28 | 42 % |
| **2_fresh** | 372 | 347 | 25 | 19 % |
| **2_old** | 372 | 347 | 25 | 22 % |

[a] The error bar was estimated to ± 2 K, corresponding to the half width at the half maximum of the fit by a Gaussian function of the magnetic susceptibility derivative.

**Table 3** Carbon, nitrogen, hydrogen and sulfur contents were determined by microanalytical procedures using an EA 1110 CHNS-O elemental analyzer from CE Instruments. Experimental elemental analysis is in good agreement with the expected calculated values.

| NPs | Elemental analysis | C [%] | N [%] | H [%] | S [%] | MW |
|---|---|---|---|---|---|---|
| **25/10** | Experimental | 26.24 | 28.82 | 3.38 | 1.66 | |
| | $[Fe(Htrz)_2(trz)_1](BF_4)\cdot(H_2O)\cdot(AOT)_{0.2}$ | 26.35 | 27.66 | 3.84 | 1.41 | 455.76 |
| **44/6** | Experimental | 29.63 | 23.13 | 3.99 | 2.54 | |
| | $[Fe(Htrz)_2(trz)_1](BF_4)\cdot(H_2O)\cdot(AOT)_{0.4}$ | 30.87 | 23.14 | 4.59 | 2.35 | 544.66 |